\documentclass[aihp]{imsart}

\RequirePackage{amsthm,amsmath,amsfonts,amssymb}
\RequirePackage[numbers]{natbib}
\RequirePackage{graphicx}

\startlocaldefs

\endlocaldefs

\begin{document}

\begin{frontmatter}

  
\title{\textbf{The Schonmann projection: how Gibbsian is it?}\\\textit{Dedicated to the memory of our dear friend Dima Ioffe}}
\runtitle{The Schonmann projection}

\thankstext{T1}{Part of the work of S.S. has been carried out at
Skoltech and at IITP RAS. The support of Russian Science Foundation (projects
No. 14-50-00150 and 20-41-09009) is gratefully acknowledged. S.S. thanks Serge
Pirogov, Yvan Velenik and Sebastien Ott for enlightening discussions. A.v.E.
thanks the participants of the Oberwolfach miniworkshop \cite{Ober}, for
various discussions as well as collaborations on related issues, and the
International Emerging Action "Long-Range" of the CNRS to make his
participation in the miniworkshop possible. He also thanks Rodrigo Bissacot,
Eric Endo and Roberto Fern\'{a}ndez for earlier collaborations and various
discussions on the issue of g-measures versus Gibbs measures. Both of the
authors are grateful to Giambattista Giacomin for his helpful comments}

\begin{aug}
\author[A]{Aernout van Enter}
\and
\author[B]{Senya Shlosman}
\address[A]{Bernoulli Institute of Mathematics,Computer Science and Artificial Intelligence\\University of Groningen, Groningen, the Netherlands\\
avanenter@gmail.com, A.C.D.van.Enter@rug.nl\}}

\address[B]{Skolkovo Institute of Science and Technology, Moscow, Russia;\\$^{\sharp}$Aix Marseille Univ, Universite de Toulon, \\CNRS, CPT, Marseille, France;\\$^{\flat}$Inst. of the Information Transmission Problems,\\RAS, Moscow, Russia. \\S.Shlosman@skoltech.ru, senya.shlosman@univ-amu.fr, shlos@iitp.ru}
\end{aug}

\begin{abstract}
We study the one-dimensional projection of the extremal Gibbs measures of the
two-dimensional Ising model -- the "Schonmann projection". These measures are
known to be non-Gibbsian at low temperatures, since their conditional
probabilities as a function of the (two-sided) boundary conditions are not
continuous. We prove the conjecture that they are g-measures, which means that
their conditional probabilities have a continuous dependence on one-sided
boundary conditions.

\end{abstract}

\begin{abstract}[language=french]
 Nous étudions la projection unidimensionnelle des mesures extrémales de Gibbs du mod\`ele d'Ising bidimensionnel - la " projection Schonmann". Ces mesures sont connues pour être non-Gibbsiennes à basses temp\'eratures, puisque leurs probabilit\'es conditionnelles en fonction
des conditions aux limites (bilatérales) ne sont pas continues. Nous prouvons la conjecture que n\'eanmoins ce sont des g-mesures, ce qui signifie que leurs
 probabilit\'es conditionnelles dépendent de façon continue des conditions aux limites unilatérales
\end{abstract}


\end{frontmatter}

\setcounter{MaxMatrixCols}{30}
\providecommand{\U}[1]{\protect\rule{.1in}{.1in}}
\newtheorem{theorem}{Theorem}
\newtheorem{acknowledgement}[theorem]{Acknowledgement}
\newtheorem{algorithm}[theorem]{Algorithm}
\newtheorem{axiom}[theorem]{Axiom}
\newtheorem{case}[theorem]{Case}
\newtheorem{claim}[theorem]{Claim}
\newtheorem{conclusion}[theorem]{Conclusion}
\newtheorem{condition}[theorem]{Condition}
\newtheorem{conjecture}[theorem]{Conjecture}
\newtheorem{corollary}[theorem]{Corollary}
\newtheorem{criterion}[theorem]{Criterion}
\newtheorem{definition}[theorem]{Definition}
\newtheorem{example}[theorem]{Example}
\newtheorem{exercise}[theorem]{Exercise}
\newtheorem{lemma}[theorem]{Lemma}
\newtheorem{notation}[theorem]{Notation}
\newtheorem{problem}[theorem]{Problem}
\newtheorem{proposition}[theorem]{Proposition}
\newtheorem{remark}[theorem]{Remark}
\newtheorem{solution}[theorem]{Solution}
\newtheorem{summary}[theorem]{Summary}

\section{Introduction}

In 1989, Roberto Schonmann published the paper: `Projections of Gibbs measures
may be non-Gibbsian', \cite{Sch}. He was considering there the Ising model on
$\mathbb{Z}^{2},$ and he showed that the projection of the $\left(  -\right)
$-phase on the $x$-axis $\mathbb{Z}^{1}\subset\mathbb{Z}^{2}$ is `too
long-range' to be qualified as a Gibbs distribution for a well-behaved
potential. In the paper \cite{DS} the situation was mended -- the definition
of the Gibbs state was properly generalized there, and from this general point
of view the Schonmann projection can still be viewed as a Gibbs measure.
Still, this projection remains an interesting object to consider.

The peculiar property of this 2D$\rightarrow$1D projection is the following.
Consider a square box $V_{M}\subset\mathbb{Z}^{2}$ of size $2M$ with $(-)$
boundary condition, where additionally we put the $(+)$ boundary condition on
two segments of the $x$-axis:
\[
I^{\prime}=[\left(  -N,0\right)  ,\left(  -n,0\right)  ],\text{ and }%
I^{\prime\prime}=[\left(  n,0\right)  ,\left(  N,0\right)  ],
\]
with $1\ll n\ll N\ll M.$ The corresponding ground state configuration
$\bar{\sigma}$ (i.e. the one which minimizes the energy) is equal to $-1$
everywhere in $V_{M}$, except on the segments $I^{\prime},$ $I^{\prime\prime
}.$ So $\bar{\sigma}$ has two contours, $\gamma^{\prime}$ and $\gamma
^{\prime\prime},$ of unit thickness, which surround the segments $I^{\prime},$
$I^{\prime\prime},$ and in particular $\bar{\sigma}_{\left(  0,0\right)
}=-1.$ However, at any positive temperature $T$ the typical configuration with
these boundary conditions looks quite different! Namely, for any $n\gg1$ and
any low temperature $T$ one can find $N\left(  n,T\right)  $ such that for any
$N>N\left(  n,T\right)  $ the typical configuration $\sigma$ looks as follows
--\ it contains just one contour $\gamma,$ surrounding both $I^{\prime}$ and
$I^{\prime\prime},$ and having $\left(  +\right)  $-phase inside. In
particular, the probability of the event $\sigma_{\left(  0,0\right)  }=+1$ is
quite large, and in fact is close to $1$. This shows that the behavior of the
projected Gibbs measure on the $x$-axis at the origin can be affected by
changing the configuration on this axis quite far away from the origin.

In this note we want to study the following question: can one similarly
influence the behavior of a typical 1D configuration $\sigma$ at the origin by
conditioning the $\left(  -\right)  $-phase to take the value $+1$ on segments
$I^{\prime},I^{\prime\prime},...$ which are allowed to be placed only
\textbf{to the left}\textit{ }of the origin. The negative answer would be an
indication that the Schonmann projection is a $g$-measure, i.e. that the
corresponding distribution is continuous with respect to the one-sided
conditioning. We remind the reader that a shift-invariant probability measure
$\mu$ on $\{-,+\}^{\mathbb{Z}}$ is called a regular $g$-measure, with $g$
being a continuous function on $\{-,+\}^{\mathbb{Z}}$, if some version of the
conditional expectation of the spin at the origin, given its complete past
$\omega$ (that is all $\omega_{i}$ with $i<0$), equals to $g$:
\[
\mu(+_{0}|\omega)-\mu(-_{0}|\omega)=g(\omega).
\]
In this paper we get this negative answer and we also prove that the
projection of the $\left(  -\right)  $-phase $\mu_{-}$ on $\mathbb{Z}^{1}$ is
a regular $g$-measure.

We will start by showing that one cannot influence the origin by one such
segment of pluses. Namely, let the $\left(  +\right)  $-segment $I^{\prime
}\subset\mathbb{Z}_{-}^{1}$ and let $\gamma^{\prime}\ $ be the shortest
possible contour surrounding $I^{\prime}$ (i.e. ground state configuration
contour). Let $T$ be some low temperature, and $\gamma$ denotes the exterior
contour of a typical configuration $\sigma,$ which surrounds the segment
$I^{\prime}$. It is well known, that the contour $\gamma$ fluctuates quite far
away from the segment $I^{\prime}$. (The phenomenon that a rigid wall repels
an interface is called entropic repulsion. It has been studied in various
contexts, and an early rigorous study can for example be found in \cite{BMF}.
For a recent one, see \cite{IST}.) However, not all fluctuations of $\gamma$
can happen! For example, the right tip of the contour $\gamma$ looks,
typically, pretty much the same as the right tip of the ground state contour
$\gamma^{\prime}.$ Moreover, the size of the fluctuations of the right tip of
$\gamma$ away from the contour $\gamma^{\prime}$ goes to zero with
temperature. Therefore, the origin stays outside the contour $\gamma,$ and in
particular the probability of the event $\sigma_{\left(  0,0\right)  }=+1$ is
very small; it is about the same as the probability of this event in the
$\left(  -\right)  $-phase.

The main technical result of the present paper is the proof of the above
picture: though the fluctuations of the contour $\gamma$ away from the $x$
axis can be quite large, they are, so to say, \textit{vertical}; the abscissa
of the rightmost tip of the contour $\gamma$ stays very close to that of
$\gamma^{\prime},$ notwithstanding the fact that the ordinate of the top of
$\gamma$ can be arbitrarily high -- it is of the order of $\sqrt{\left\vert
I^{\prime}\right\vert }$. In other words, although the transversal
fluctuations diverge, the longitudinal ones remain bounded uniformly in $n$
and $N$. Our proof uses cluster expansion and so is limited to low
temperatures $T$, though we believe the result to be true for all $T<T_{cr}.$

In the next section we will describe our general strategy. We will then
provide the details in the third section. Then we explain how this technical
result implies $g$-measurability.

\section{Main technical result}

We are considering the 2-dimensional nearest-neighbour ferromagnetic Ising
model, with formal Hamiltonian $-H=\sum_{i,j\in\mathbb{Z}^{2}}\sigma_{i}%
\sigma_{j}$, at low temperature. It is known that there exist two extremal
Gibbs measures for this model, $\mu^{+}$ (the maximal one) and $\mu^{-}$ (the
minimal one ), which are shift-invariant probability measures on
${\{-,+\}}^{\mathbb{Z}^{2}}$. See e.g. \cite{Geo} and/or \cite{FV}\textbf{ }
for this and further background on Gibbs measure theory.

Our main result is about the one-dimensional marginal (the Schonmann
projection) of the measures above. As mentioned before, these measures are
known not to be Gibbs measures.

Our analysis will make extensive use of contours, which are curves made of
bonds in the dual lattice of $\mathbb{Z}^{2}$, separating plus and minus
spins. Contours have been a major tool in understanding the low-temperature
Ising model since Peierls introduced them in the 1930's. Studying measures on
contour configurations is equivalent to studying measures on spin
configurations. In particular, at low temperatures, long contours tend to be
exponentially improbable, except in cases when they are unavoidable, which can
happen due to a specific choice of boundary conditions.

We want to show that the spin at the origin depends in a continuous manner on
the spins in its one-dimensional past, that is, on the spins on the $x$-axis
to the left of the origin.

We concentrate on the low\textbf{-}$T$ 2D Ising model in a square box $V_{M}$
of size $2M$ with $(-)$ boundary condition, where additionally we put the
$(+)$ boundary condition on a segment $I=[\left(  -N,0\right)  ,\left(
0,0\right)  ],N<M,$ thereby forcing a contour $\gamma$ around $I$. We are
interested in the probability that the point $\left(  n,0\right)  \in
$\textrm{Int}$\left(  \gamma\right)  ;$ we want to show that it is of the
order $\exp\left\{  -\beta n\right\}  ,$ uniformly in $N$ and $M.$

We note that this result concerns the situation when the spins in the interval
between $\left(  0,0\right)  $ and $\left(  n,0\right)  $ are not fixed. This
then implies that \textbf{for most} fixed choices of spin configurations in
this interval the dependence on what happens outside it decays exponentially,
as was proven in \cite{BC}. \newline In section 4 we will explain that in fact
the dependence decays \textbf{for all} fixed choices of spin configurations in
this interval, but for some choices only algebraically. This decay property
will prove the $g$-measure property of the\textbf{ } Schonmann projection.

\begin{theorem}
The probability of the event that the point $\left(  n,0\right)  \ $happens to
be inside the contour $\gamma$ is of the order $\exp\left\{  -\beta n\right\}
,$ uniformly in $N$ and $M,$ provided $\beta$ is large enough.
\end{theorem}

This claim is not at all immediate, due to the entropic repulsion of $\gamma$
from $I.$ If the segment $I$ is very long, then the contour $\gamma$ would go
away from $I$ by a distance $\sim\sqrt{\left\vert I\right\vert }.$ Yet, this
deviation of $\gamma$ from $I$ goes in vertical direction only, and the
contour $\gamma$ passes quite close to the edges $\left(  -N,0\right)  ,$
$\left(  0,0\right)  $ of the segment, as we are going to show.

So let $\gamma$ be our (exterior) contour. Its distribution is given by the
weight%
\begin{equation}
w\left(  \gamma\right)  =\exp\left\{  -\beta\left\vert \gamma\right\vert
+\sum_{\Lambda:\Lambda\cap\gamma\neq\varnothing}\Phi\left(  \Lambda\right)
\right\}  , \label{02}%
\end{equation}
where $\left\vert \gamma\right\vert $ is the length of the contour $\gamma,$
the summation over $\Lambda$ goes over connected subsets $\Lambda\subset
V_{M}\setminus I,$ and $\Phi\left(  \Lambda\right)  $-s are exponentially
small in $\mathrm{diam}\left(  \Lambda\right)  :$
\begin{equation}
\left\vert \Phi\left(  \Lambda\right)  \right\vert \leq\exp\left\{
-2\beta\mathrm{diam}\left(  \Lambda\right)  \right\}  . \label{13}%
\end{equation}
Such a representation is well-known; one can find it in \cite{DKS}, where the
function $\Phi$ is also explicitly defined.

Let $L=\left\{  \left(  x,y\right)  :x=-N\right\}  $, $R=\left\{  \left(
x,y\right)  :x=0\right\}  $ be two vertical lines, $S$ be the strip between
them, and $H_{\pm}$ be the upper and lower half-planes. Define the set of four
cut-points $u_{1},u_{2},v_{1},v_{2}$ of $\gamma$ by the properties:

\begin{itemize}
\item $u_{1},u_{2}\in\gamma\cap L,$ $v_{1},v_{2}\in\gamma\cap R,$

\item the arcs $\gamma_{1}$ (piece of $\gamma$ from $u_{1}$ to $v_{1}$) and
$\gamma_{2}$ (piece of $\gamma$ from $u_{2}$ to $v_{2}$) lie inside the strip
$S.$
\end{itemize}

We denote by $\gamma_{u},\gamma_{v}$ the remaining two arcs of $\gamma.$

Let us define
\[
Z\left(  u_{1},u_{2},v_{1},v_{2}\right)  =\sum_{\substack{\gamma:\\u_{1}%
,u_{2},v_{1},v_{2}\in\gamma}}w\left(  \gamma\right)  ,
\]
where the summation goes over $\gamma$-s with cut-points $u_{1},u_{2}%
,v_{1},v_{2}.$

We first pretend that
\[
w\left(  \gamma\right)  =w\left(  \gamma_{u}\right)  w\left(  \gamma
_{1}\right)  w\left(  \gamma_{v}\right)  w\left(  \gamma_{2}\right)  ,
\]
and moreover%
\begin{equation}
Z\left(  u_{1},u_{2},v_{1},v_{2}\right)  \approx\exp\left\{  -\beta\left(
u_{1}-u_{2}+v_{1}-v_{2}\right)  \right\}  Z\left(  u_{1}\rightarrow
v_{1}\right)  Z\left(  u_{2}\rightarrow v_{2}\right)  , \label{10}%
\end{equation}
where $Z\left(  u_{1}\rightarrow v_{1}\right)  =\sum_{\gamma_{1}}w\left(
\gamma_{1}\right)  ,$ $Z\left(  u_{2}\rightarrow v_{2}\right)  =\sum
_{\gamma_{2}}w\left(  \gamma_{2}\right)  ,$ and the weights $w\left(
\gamma_{1}\right)  ,w\left(  \gamma_{2}\right)  $ are taken from $\left(
\ref{02}\right)  ,$ i.e.%
\begin{equation}
w\left(  \gamma_{1}\right)  =\exp\left\{  -\beta\left\vert \gamma
_{1}\right\vert +\sum_{\Lambda:\Lambda\cap\gamma_{1}\neq\varnothing}%
\Phi\left(  \Lambda\right)  \right\}  , \label{03}%
\end{equation}
where now both $\gamma_{1}$ and all $\Lambda$-s stay in the upper semistrip
$I\times\{1,2,3,...\},$ and similarly for $w\left(  \gamma_{2}\right)  .$
(Compare with $\left(  \ref{02}\right)  $.)

In what follows we keep $N$ fixed and we abuse notation by writing $u_{1}$
instead of $\left(  -N,u_{1}\right)  ,$ etc.

While considering the partition functions $Z\left(  u_{1}\rightarrow
v_{1}\right)  $, $Z\left(  u_{2}\rightarrow v_{2}\right)  $ we will separate
between regimes when the points $u_{i},v_{i}$ are of order $\sqrt{N}$ or
smaller, or are of higher orders.

\begin{enumerate}
\item Suppose first that $u_{1}>\sqrt{N},$ and $v_{1}>u_{1}.$ Then,
\[
Z\left(  u_{1}\rightarrow v_{1}\right)  <Z\left(  u_{1}\rightarrow
u_{1}\right)
\]
(meaning: $Z\left(  u_{1}\rightarrow u_{1}\right)  \equiv Z\left(  \left(
-N,u_{1}\right)  \rightarrow\left(  0,u_{1}\right)  \right)  $). Moreover, the
function $Z\left(  u_{1}\rightarrow v_{1}\right)  $ is decreasing in $v_{1}$
in this regime. Because of the extra factor of $\exp\left\{  -\beta\left(
v_{1}-u_{1}\right)  \right\}  ,$ coming from the weight $w\left(  \gamma
_{v}\right)  ,$ we can disregard the contribution of the configurations with
$u_{1}>\sqrt{N},$ and $v_{1}>u_{1}$ to the partition function $\sum w\left(
u_{1},u_{2},v_{1},v_{2}\right)  .$

\item The same monotonicity in $v_{1}$ holds when $u_{1}\leq\sqrt{N},$ and
$v_{1}>\sqrt{N},$ with the same conclusion.

\item Iterating 1, 2,\textbf{ }we come to the remaining case, when all four
variables $u_{1},u_{2},v_{1},v_{2}$ are $\leq\sqrt{N}$ in absolute values. In
this regime we use the local limit theorem for the random point $v$ in the
ensemble $Z\left(  u_{1}\rightarrow v\right)  $ and claim the existence of a
constant $C,$ which does not depend on $\beta,$ $u$ and $v,$ such that%
\begin{equation}
\frac{Z\left(  u_{1}\rightarrow\left(  v_{1}+1\right)  \right)  }{Z\left(
u_{1}\rightarrow v_{1}\right)  }<C. \label{01}%
\end{equation}
For $\beta$ large, the factor $\exp\left\{  -\beta\right\}  $ -- which is the
price, due to the weight $w\left(  \gamma_{v}\right)  ,$ for the point $v_{1}$
to be one step higher -- beats $C.$
\end{enumerate}

\section{Splitting the partition function}

We start by presenting the rigorous counterpart to the relation $\left(
\ref{10}\right)  .$

We will use the standard SW convention to define the contours as self-avoiding
loops. Such contours and self-avoiding lattice paths will be called legal. Let
$\gamma$ be an exterior contour surrounding the segment $\left[  \left(
-N,0\right)  ,\left(  0,0\right)  \right]  .$ Then the cut-point $u_{1}$ is
defined as the point of the last intersection of $\gamma$ with the line $L$
before getting to the line $R,$ provided $\gamma$ is oriented clockwise. The
other three points $u_{2},v_{1},v_{2}$ are defined similarly.

As a result, $\gamma$ is a concatenation,
\[
\gamma=\gamma_{1}\circ\gamma_{v}\circ\gamma_{2}\circ\gamma_{u},
\]
where

\begin{itemize}
\item $\gamma_{1}$ is any legal path in $S,$ joining $u_{1}$ and $v_{1}$ and
lying above $\left[  \left(  -N,0\right)  ,\left(  0,0\right)  \right]  ,$

\item $\gamma_{2}$ is any legal path in $S,$ joining $u_{2}$ and $v_{2}$ and
lying below $\left[  \left(  -N,0\right)  ,\left(  0,0\right)  \right]  ,$

\item $\gamma_{u}$ $\ $(resp. $\gamma_{v}$) is any legal path joining $u_{1}$
and $u_{2}$ (joining $v_{1}$ and $v_{2}$), not intersecting $\left[  \left(
-N,0\right)  ,\left(  0,0\right)  \right]  $, and such that the concatenation
$\gamma_{1}\circ\gamma_{v}\circ\gamma_{2}\circ\gamma_{u}$ is legal.
\end{itemize}

\noindent As this definition suggests, we will treat the arcs $\gamma
_{1},\gamma_{2}$ of $\gamma$ as independent variables, which put restrictions
on the allowed realizations of $\gamma_{u},\gamma_{v}$.

Having in mind the definition $\left(  \ref{02}\right)  ,$ we write%

\begin{align}
&  Z\left(  u_{1},u_{2},v_{1},v_{2}\right) \nonumber\\
&  =\sum_{\substack{\gamma_{1}:u_{1}\rightarrow v_{1}\\\gamma_{2}%
:u_{2}\rightarrow v_{2}}}\left[  \exp\left\{  -\beta\left\vert \gamma
_{1}\right\vert +\sum_{\substack{\Lambda:\Lambda\cap\gamma_{1}\neq
\varnothing,\\\Lambda\subset H_{+}}}\Phi\left(  \Lambda\right)  \right\}
\exp\left\{  -\beta\left\vert \gamma_{2}\right\vert +\sum_{\substack{\Lambda
:\Lambda\cap\gamma_{2}\neq\varnothing,\\\Lambda\subset H_{-}}}\Phi\left(
\Lambda\right)  \right\}  \right. \label{12}\\
&  \times\left(  \sum_{_{\substack{\gamma_{u}:u_{1}\rightarrow u_{2}%
,\gamma_{v}:v_{1}\rightarrow v_{2}\\\gamma_{1}\circ\gamma_{v}\circ\gamma
_{2}\circ\gamma_{u}\text{ is legal}}}}\exp\left\{  -\beta\left\vert \gamma
_{u}\right\vert +\sum_{\substack{\Lambda:\Lambda\cap\gamma_{u}\neq
\varnothing,\\\Lambda\cap\left(  \gamma_{1}\cup\gamma_{2}\right)
=\varnothing}}\Phi\left(  \Lambda\right)  \right\}  \right. \nonumber\\
&  \left.  \left.  \exp\left\{  -\beta\left\vert \gamma_{v}\right\vert
+\sum_{\substack{\Lambda:\Lambda\cap\gamma_{v}\neq\varnothing,\\\Lambda
\cap\left(  \gamma_{1}\cup\gamma_{2}\cup\gamma_{u}\right)  =\varnothing_{-}%
}}\Phi\left(  \Lambda\right)  \right\}  \right)  \right] \nonumber
\end{align}

\subsection{The vertical parts \label{exp}}

Let us first consider the partition function
\begin{equation}
Z\left(  v_{1},v_{2}\right)  =\sum_{_{\gamma:v_{1}\rightarrow v_{2}}}%
\exp\left\{  -\beta\left\vert \gamma\right\vert +\sum_{\Lambda:\Lambda
\cap\gamma\neq\varnothing,}\Phi\left(  \Lambda\right)  \right\}  \label{11}%
\end{equation}
(where we do not have the restriction that $\gamma$ stays away from the
segment $I.$) It is well-known and easy to show that for any $\varepsilon>0$
the probability of the event that $\left\vert \gamma\right\vert >\left(
1+\varepsilon\right)  \left\Vert v_{1}-v_{2}\right\Vert $ goes to zero as
$\beta\rightarrow\infty,$ uniformly in $v_{1},v_{2}.$ (For the benefit of the
reader we will outline the proof of this statement at the end of this
subsection.) Therefore, the probability $p_{1}\left(  \beta\right)  $ in the
ensemble $\left(  \ref{11}\right)  $ of the event that the `first' edge of
$\gamma$ -- i.e. the edge starting from $v_{1}$ -- goes down in the direction
$v_{2},$ thus connecting $v_{1}$ to $\left(  v_{1}-e_{2}\right)  $, has the
property that $p_{1}\left(  \beta\right)  \rightarrow1$ as $\beta
\rightarrow\infty.$ Let $Z^{\downarrow}\left(  v_{1},v_{2}\right)  $ be part
of the partition function $\left(  \ref{11}\right)  $ restricted to such
configurations. Clearly,
\[
\frac{Z^{\downarrow}\left(  v_{1},v_{2}\right)  }{Z\left(  v_{1}-e_{2}%
,v_{2}\right)  }\sim e^{-\beta},
\]
since the terms $\Phi\left(  \Lambda\right)  $ are of smaller order.
Therefore,
\[
\frac{Z\left(  v_{1},v_{2}\right)  }{Z\left(  v_{1}-e_{2},v_{2}\right)
}=\frac{1}{p_{1}\left(  \beta\right)  }\frac{Z^{\downarrow}\left(  v_{1}%
,v_{2}\right)  }{Z\left(  v_{1}-e_{2},v_{2}\right)  }\leq e^{-c\beta},
\]
for some $c\rightarrow1$ as $\beta\rightarrow\infty.$ The same argument,
slightly modified, applies to the third and the fourth partition functions in
$\left(  \ref{12}\right)  .$

\begin{lemma}
Let $v_{1},v_{2}\in\mathbb{Z}^{2}$ be two points on the lattice. Consider the
ensemble $P$ of lattice paths $\gamma,$ joining these two points, and defined
by the weight $w\left(  \gamma\right)  =\exp\left\{  -\beta\left\vert
\gamma\right\vert \right\}  .$ Let $\left\Vert v_{1}-v_{2}\right\Vert $ be the
length of the shortest such path. Let $\varepsilon>0,$ and consider the event
$\left\{  \gamma:\left\vert \gamma\right\vert >\left(  1+\varepsilon\right)
\left\Vert v_{1}-v_{2}\right\Vert \right\}  .$ Then $P\left\{  \gamma
:\left\vert \gamma\right\vert >\left(  1+\varepsilon\right)  \left\Vert
v_{1}-v_{2}\right\Vert \right\}  \rightarrow0$ as $\beta\rightarrow\infty,$
uniformly in $v_{1},v_{2}.$
\end{lemma}

\begin{proof}
Let $\bar{\gamma}:v_{1}\rightarrow v_{2}$ be one of the paths with $\left\vert
\bar{\gamma}\right\vert =\left\Vert v_{1}-v_{2}\right\Vert .$ Then%
\begin{align*}
&  P\left\{  \gamma:\left\vert \gamma\right\vert >\left(  1+\varepsilon
\right)  \left\Vert v_{1}-v_{2}\right\Vert \right\} \\
&  \leq\frac{\sum_{l>\left(  1+\varepsilon\right)  \left\Vert v_{1}%
-v_{2}\right\Vert }3^{l}\exp\left\{  -\beta l\right\}  }{\exp\left\{
-\beta\left\vert \bar{\gamma}\right\vert \right\}  }\\
&  =\frac{\sum_{k>0}3^{\left(  1+\varepsilon\right)  \left\Vert v_{1}%
-v_{2}\right\Vert +k}\exp\left\{  -\beta\left[  \left(  1+\varepsilon\right)
\left\Vert v_{1}-v_{2}\right\Vert +k\right]  \right\}  }{\exp\left\{
-\beta\left\Vert v_{1}-v_{2}\right\Vert \right\}  }\\
&  =\sum_{k>0}3^{\left(  1+\varepsilon\right)  \left\Vert v_{1}-v_{2}%
\right\Vert +k}\exp\left\{  -\beta\left[  \varepsilon\left\Vert v_{1}%
-v_{2}\right\Vert +k\right]  \right\}  \rightarrow0
\end{align*}
as $\beta\rightarrow\infty,$ since the factor $\exp\left\{  -\beta
\varepsilon\left\Vert v_{1}-v_{2}\right\Vert \right\}  $ beats $3^{\left(
1+\varepsilon\right)  \left\Vert v_{1}-v_{2}\right\Vert }$ once $\beta$ is
large enough (depending on $\varepsilon$), while $\exp\left\{  -\beta
k\right\}  $ beats $3^{k}.$
\end{proof}

\subsection{The horizontal parts \label{f51}}

Here we will treat the partition function%
\[
Z\left(  u_{1}\rightarrow v_{1}\right)  =\sum_{\gamma_{1}:u_{1}\rightarrow
v_{1}}\exp\left\{  -\beta\left\vert \gamma_{1}\right\vert +\sum
_{\substack{\Lambda:\Lambda\cap\gamma_{1}\neq\varnothing,\\\Lambda\subset
H_{+}}}\Phi\left(  \Lambda\right)  \right\}  .
\]
The properties needed are obtained in \cite{IOVW}, which is based on the
random walk approximation of the random line $\gamma_{1}$, worked out in
\cite{OV}. One can use instead the random walk description introduced in
\cite{DS}, and used for a similar goal in \cite{IST}.

In this subsection we will drop the subscript $1$ and will write $u,v,\gamma$
instead of $u_{1},v_{1},\gamma_{1}.$

The model we have to deal with is defined by assigning the weight
\[
w\left(  \gamma\right)  =\exp\left\{  -\beta\left\vert \gamma\right\vert
+\sum_{\substack{\Lambda:\Lambda\cap\gamma\neq\varnothing,\\\Lambda\subset
H_{+}}}\Phi\left(  \Lambda\right)  \right\}
\]
to any path $\gamma\subset S.$ As in \cite{IST}, we can pass to an enlarged
ensemble, with more variables -- $\left(  \gamma,\mathbf{\Lambda}\right)  $ --
consisting of a path $\gamma$ and a finite collection $\mathbf{\Lambda}$ of
connected sets, $\mathbf{\Lambda=}\left\{  \Lambda_{i}\subset\mathbb{Z}%
^{2}\right\}  ,$ \textit{each intersecting} $\gamma,$ and defined by the
weight%
\begin{equation}
\mathbf{w}\left(  \gamma,\mathbf{\Lambda}\right)  =\exp\left\{  -\beta
^{\prime}\left\vert \gamma\right\vert \right\}  \prod_{\Lambda_{i}%
\in\mathbf{\Lambda}}\Psi\left(  \Lambda_{i}\right)  . \label{15}%
\end{equation}
The special case of $\mathbf{\Lambda=\varnothing}$ is not excluded. Here the
functional $\Psi$ satisfies the same estimate $\left(  \ref{13}\right)  ,$ but
in addition is \textit{positive}, which makes $\mathbf{w}$ a legitimate
statistical weight. The functional $\Psi$ and the new temperature
$\beta^{\prime}$ can be chosen in such a way that $\sum_{\mathbf{\Lambda}%
}\mathbf{w}\left(  \gamma,\mathbf{\Lambda}\right)  =w\left(  \gamma\right)  $
-- so the partition functions for the weights $\mathbf{w}$ and $w$ are the
same -- while $\left\vert \beta-\beta^{\prime}\right\vert \rightarrow0$ as
$\beta\rightarrow\infty.$ We will call a pair $\left(  \gamma,\mathbf{\Lambda
}\right)  $ a \textit{dressed} path, or just a path. The idea of introducing
the hidden variables $\mathbf{\Lambda}$ to the ensemble $\left(
\gamma,\mathbf{\Lambda}\right)  $ goes back to \cite{DS}, see \cite{IST} for
more details.

Let $x_{0}\in\mathbb{Z}^{1}.$ We call the point $x_{0}$ a splitting point of
the dressed path $\left(  \gamma,\mathbf{\Lambda}\right)  ,$ if the
intersection of the line $l_{x_{0}}=\left\{  \left(  x,y\right)
:x=x_{0}\right\}  $ with the curve $\gamma$ is a single point, while all the
intersections $l_{x_{0}}\cap\Lambda_{i}=\varnothing,$ $\Lambda_{i}%
\in\mathbf{\Lambda.}$ Let $x_{1},...,x_{k}$ be all the splitting points of the
path $\left(  \gamma,\mathbf{\Lambda}\right)  .$ Then $\left(  \gamma
,\mathbf{\Lambda}\right)  $ is split by the lines $l_{x_{i}}$ into $k+1$
irreducible pieces $\left(  \gamma_{0},\mathbf{\Lambda}_{0}\right)
,...,\left(  \gamma_{k},\mathbf{\Lambda}_{k}\right)  ,$ and the dressed path
$\left(  \gamma,\mathbf{\Lambda}\right)  $ is their concatenation. We will
call the irreducible pieces $\left(  \gamma_{i},\mathbf{\Lambda}_{i}\right)  $
the \textit{animals. }Note that%
\[
\mathbf{w}\left(  \gamma,\mathbf{\Lambda}\right)  =\prod_{i=0}^{k}%
\mathbf{w}\left(  \gamma_{i},\Lambda_{i}\right)  ,
\]
which paves the way to the definition of the random walk $\mathsf{S}$ -- the
effective random walk representation.

Let $u=\left(  x,y\right)  ,u^{\prime}=\left(  x^{\prime},y^{\prime}\right)  $
be two points in $\mathbb{Z}^{2},$ $x<x^{\prime}.$ We define the weight
\[
s_{u,u^{\prime}}=\sum_{\substack{\left(  \gamma,\mathbf{\Lambda}\right)
:\\\gamma:u\rightarrow u^{\prime}}}^{\prime}\mathbf{w}\left(  \gamma
,\mathbf{\Lambda}\right)  ,
\]
where the summation goes over dressed paths $\left(  \gamma,\mathbf{\Lambda
}\right)  $ such that

\begin{itemize}
\item the path $\gamma$ goes from $u$ to $u^{\prime},$ and

\item the dressed path $\left(  \gamma,\mathbf{\Lambda}\right)  $ is its
unique irreducible piece.
\end{itemize}

\noindent These weights define the distribution of the random vector
$X=\left(  \theta,\zeta\right)  ,$ which defines the steps of the walk
$\mathsf{S}$. Its starting point will be $\mathsf{S}_{0}=\left(  -N,u\right)
,$ while $\mathsf{S}_{i}=\left(  -N,u\right)  +\sum_{j=1}^{i}X_{j}%
=u_{1}+\left(  \mathsf{T}_{i},\mathsf{Z}_{i}\right)  ,$ where $\mathsf{T}%
_{i}=\sum_{j}\theta_{j},$ $\mathsf{Z}_{i}=\sum_{j}\zeta_{j},$ (we follow here
the notations of \cite{IOVW}, definition (45) ). The overall distribution of
$\mathsf{S}$ will be denoted by $\mathbf{P}_{\left(  -N,u\right)  }.$

To study the ratio $\frac{Z\left(  u\rightarrow\left(  v+1\right)  \right)
}{Z\left(  u\rightarrow v\right)  }$ in $\left(  \ref{01}\right)  $ we can
pass to the study of the probabilities in the ensemble $\mathbf{P}_{\left(
-N,u\right)  }$ of the event $\left\{  \mathsf{S}:\left(  -N,u\right)
\rightarrow\left(  0,v\right)  ,\mathsf{S}>0\right\}  $ that the path
$\mathsf{S}$ stays positive and arrives to the point $\left(  0,v\right)  ,$
resp. $\left(  0,v+1\right)  .$ The very precise estimates of \cite{IOVW}, see
the relations (47-49) there, tell us that
\begin{equation}
\mathbf{P}_{\left(  -N,u\right)  }\left\{  \mathsf{S}:\left(  -N,u\right)
\rightarrow\left(  0,v\right)  ;\mathsf{S}>0\right\}  \sim C\frac{h^{+}\left(
u\right)  h^{-}\left(  v\right)  }{N^{3/2}} \label{51}%
\end{equation}
as $N\rightarrow\infty,$ where

\begin{itemize}
\item the function $h^{+}\left(  x\right)  =x-\mathbb{E}_{x}\left(
\mathsf{Z}_{\tau}\right)  ,$ where the random walk $\mathsf{Z}$ starts from
the point $x\in\mathbb{Z}^{1},$ $x>0,$ and the stopping moment $\tau$ is
defined by $\tau=\inf\left\{  n:\mathsf{Z}_{n}\leq0\right\}  ;$

\item the function $h^{-}\left(  x\right)  $ is defined in the same way, but
for the random walk $\left(  -\mathsf{Z}\right)  ;$

\item $C=C\left(  \theta,\zeta\right)  >0$ is some constant;

\item the variables $u,v\in\left[  1,N^{1/2-\delta}\right]  ,$ with any small
$\delta>0,$ which parameter will be fixed from now on (say, $\delta=\frac
{1}{100}$).
\end{itemize}

For the region $u\in\left[  1,N^{1/2-\delta}\right]  ,$ $v\in\left[
N^{1/2-\delta},N^{1/2}\right]  $ we have
\begin{equation}
\mathbf{P}_{\left(  -N,u\right)  }\left\{  \mathsf{S}:\left(  -N,u\right)
\rightarrow\left(  0,v\right)  ;\mathsf{S}>0\right\}  \sim C\frac{h^{+}\left(
u\right)  v\exp\left\{  -v^{2}/2N\right\}  \mathbf{Var}\left(  \zeta\right)
}{N^{3/2}}, \label{52}%
\end{equation}
while in the region $u,v\in\left[  N^{1/2-\delta},N^{1/2}\right]  $ we have%
\begin{equation}
\mathbf{P}_{\left(  -N,u\right)  }\left\{  \mathsf{S}:\left(  -N,u\right)
\rightarrow\left(  0,v\right)  ;\mathsf{S}>0\right\}  \sim\frac{\psi\left(
u/N^{1/2},v/N^{1/2}\right)  }{N^{1/2}} \label{53}%
\end{equation}
for some positive continuous bounded function $\psi$ on $\left[  0,1\right]
^{2}.$

Since in our case the random variable $\zeta$ is exponentially localized, i.e.
$\Pr\left\{  \zeta=k\right\}  \sim\exp\left\{  -c\left(  \beta\right)
\left\vert k\right\vert \right\}  $ with $c\left(  \beta\right)
\rightarrow\infty$ as $\beta\rightarrow\infty,$ we have that
\[
\frac{\mathbf{P}_{\left(  -N,u\right)  }\left\{  \mathsf{S}:\left(
-N,u\right)  \rightarrow\left(  0,v+1\right)  ;\mathsf{S}>0\right\}
}{\mathbf{P}_{\left(  -N,u\right)  }\left\{  \mathsf{S}:\left(  -N,u\right)
\rightarrow\left(  0,v\right)  ;\mathsf{S}>0\right\}  }<C
\]
uniformly for all $N$ large enough and $u,v\in\left[  1,N^{1/2}\right]  .$ In
words, though the probability $\mathbf{P}_{\left(  -N,u\right)  }$ of the
event $\left\{  \mathsf{S}:\left(  -N,u\right)  \rightarrow\left(
0,v+1\right)  ;\mathsf{S}>0\right\}  $ can be higher than $\mathbf{P}_{\left(
-N,u\right)  }\left\{  \mathsf{S}:\left(  -N,u\right)  \rightarrow\left(
0,v\right)  ;\mathsf{S}>0\right\}  $ due to the entropic repulsion of the path
$\mathsf{S}$ from the $x$-axis, their ratio is bounded by a constant. That
proves $\left(  \ref{01}\right)  .$

Combining the horizontal and vertical parts implies that with large
probability the cut points $u_{1},u_{2},v_{1}$ and $v_{2}$ all are at a not
too large vertical distance of the segment, which distance does not grow with
$N$. Therefore the probability that a point at horizontal distance $n$ to the
right of the segment will be inside the contour containing the segment decays
exponentially in $n$, again uniformly in $N$ (and $M$). For the projected
system this implies the continuity of the spin expectation in the origin on
our left configuration.

\section{Uniform continuity}

Here we prove the main result of our paper.

\begin{theorem}
The Schonmann projection, at sufficiently low temperature, is a regular g-measure
\end{theorem}

\textbf{Proof. }Let $\omega$ be some semi-infinite string of spins
$...\sigma_{-N-1},\sigma_{-N},\sigma_{-N+1},...,\sigma_{-1}.$ Let us introduce
the notation $\omega_{N+}$ for the configuration $...+,+,+,\sigma_{-N}%
,\sigma_{-N+1},...,\sigma_{-1},$ and denote by $\omega_{N-}$ the configuration
$...-,-,-,\sigma_{-N},\sigma_{-N+1},...,\sigma_{-1}.$ To prove the uniform
continuity of our $g$-function we will look at the difference
\[
\mu(+_{0}|\omega_{N+})-\mu(+_{0}|\omega_{N-})\equiv g^{\prime}(\omega,N)\geq0.
\]
We will show that $g^{\prime}(\omega,N)\rightarrow0$ as $N\rightarrow\infty,$
uniformly in $\omega.$ Clearly, this implies the uniform continuity, because
of the FKG property.

Depending on the context, we will consider below the quantities $\mu
_{M^{\prime}}(+_{0}|\omega_{M^{\prime\prime}}),$ where $M^{\prime}\geq
M^{\prime\prime},$ $\mu_{M^{\prime}}$ is the Gibbs distribution in the box
$V_{M^{\prime}}$ with $\left(  -\right)  $-boundary condition, and
$\omega_{M^{\prime\prime}}$ is a configuration on the segment $\left[
-M^{\prime\prime},1\right]  .$ To save on notation\textbf{, } we sometime will
denote them by the same expression $\mu(+_{0}|\omega).$

\textbf{1. }The simplest case to consider is the string $\omega=\omega
^{+}\equiv+1.$ This case is basically the one considered above. We need to
compute the difference\newline%
\[
\mu_{M}(+_{0}|\omega_{M}^{+})-\mu_{M}(+_{0}|\omega_{N-}^{+})=\frac{\mu
_{M}(+_{0},\omega_{M}^{+})}{\mu_{M}(\omega_{M}^{+})}-\frac{\mu_{M}%
(+_{0},\omega_{N-}^{+})}{\mu_{M}(\omega_{N-}^{+})},
\]
where the size of the box, $M$, exceeds the length $N$ of the segment. The
ratio of two probabilities, $\frac{\mu_{M}(+_{0},\omega_{N-}^{+})}{\mu
_{M}(\omega_{N-}^{+})},$ is the ratio of the two partition functions: one is
taken over all configurations containing a contour $\gamma$ surrounding the
segment $\left[  -N,-1\right]  ,$ while the other is restricted to those
$\gamma$-s, which enclose an extra point $\left(  0,0\right)  .$ According to
sections \ref{exp}, \ref{f51}, this ratio is equivalent to
\[
\exp\left\{  -2\beta\right\}  \left(  \frac{N^{3/2}}{\left(  N+1\right)
^{3/2}}\right)  ^{2},
\]
see $\left(  \ref{12},\ref{51}-\ref{53}\right)  .$ Indeed, the loop $\gamma$
surrounding the segment $\left[  -N,-1\right]  $ roughly corresponds to a pair
of paths (one above and one below the segment), and each of them has to pass
right near the tip of the segment. (The square appears here because $\gamma$
contains two horizontal strings.) The same analysis applies to the term
$\frac{\mu_{M}(+_{0},\omega_{M}^{+})}{\mu_{M}(\omega_{M}^{+})},$ with $N$
replaced by $M>N.$  So their difference $g^{\prime}(\omega^{+},N)\lesssim
\frac{1}{N}.$

\textbf{2. }Now consider the string $\omega^{-}\equiv-1.$ Here we need to know
the behavior of the contour $\gamma,$ surrounding the $\left(  +\right)
$-segment $\left[  -M,-M+N\right]  ,$ and we assume that $M>N.$ The contour
$\gamma$ must cross the $x$-axis to the right of $\left[  -M,-M+N\right]  ,$
and it has two options to do so: one is to cross at the location $-N;$ the
other is to cross through a point at a positive semiaxis.

In the first case, the contour $\gamma$ is at distance $N$ from the origin, so
the contribution to $g^{\prime}(\omega^{-},N)$ is exponentially small in $N,$
as follows from the cluster expansion.

In the second case the $\left(  -\right)  $-segment $\left[  -N,-1\right]  $
gets inside the $\left(  +\right)  $-phase, which fills the interior
\textrm{Int}$\left(  \gamma\right)  $ of the contour $\gamma,$ so it is
surrounded by an extra contour $\Gamma\subset$\textrm{Int}$\left(
\gamma\right)  $, which brings an extra cost of $\exp\left\{  -2\beta
\left\vert \Gamma\right\vert \right\}  .$ Note that under condition
$\Gamma\subset$\textrm{Int}$\left(  \gamma\right)  $ the behavior of $\gamma$
can be quite different from the one we have seen above; in particular the
event that $\left(  0,0\right)  \in$\textrm{Int}$\left(  \gamma\right)  $ can
be quite likely, in which case the magnetization at the origin is very
different from $-m^{\ast}\left(  \beta\right)  .$ However, the factor
$\exp\left\{  -4\beta N\right\}  $ -- which is the probability of the
appearance of $\Gamma$ -- beats all these complications, and we conclude that
$g^{\prime}(\omega^{-},N)=o\left(  \exp\left\{  -\beta N\right\}  \right)  .$

\textbf{3.} Next, consider the key (and most delicate) case of the string
$\omega^{-,k}=\left(  ...+++\underset{k}{\underbrace{--...-}}\right)  ,$ with
$k$ $\left(  -\right)  $-spins neighboring the origin, $k=1,2,...$ . Again,
the contour $\gamma,$ surrounding the $\left(  +\right)  $-string, can cross
the $x$-axis either at the location $-k$ or at some location to the right of
the point $\left(  -1,0\right)  .$ In the second case we again have the
contour $\Gamma\subset$\textrm{Int}$\left(  \gamma\right)  .$ But this time
the small weight $\exp\left\{  -2\beta\left\vert \Gamma\right\vert \right\}  $
is small in $k$ only -- but not in $N$, so it gives us the estimate
$g^{\prime}(\omega^{-k},N)=o\left(  \exp\left\{  -\beta k\right\}  \right)  ,$
which is not what we look for, since we need the estimate decaying in $N$.

\textbf{3a. }Anyway, the analysis of the first case, when the contour $\gamma$
crosses the $x$-axis at the location $-k$ is similar to the case \textbf{1},
and we conclude that its contribution to the function $g^{\prime}(\omega
^{-,k},N)\sim\frac{c\left(  k\right)  }{N}$ with $c\left(  k\right)
\rightarrow0$ as $k\rightarrow\infty,$ which is more than enough for our purposes.

\textbf{3b. }As for the second case, the presence of the contour $\Gamma$ can
change the magnetization at the origin quite a bit. Our point now is that
\textbf{the change is almost the same for both cases} -- when $\gamma$
surrounds the segment $\left[  -N,-1\right]  $ and when it surrounds the
segment $\left[  -M,-1\right]  $ with $M>N,$ their difference being $\sim
\frac{1}{N}.$

To see it, let us fix some scale, growing with $N$ -- say, $\ln N$ -- and
consider two cases. The first is when $\left\vert \Gamma\right\vert >\ln N.$
Then the probability of the appearance of such a $\Gamma$ decays as
$N^{-\beta},$ so what happens with the spin at the origin due to the contour
$\gamma$ under the condition $\Gamma\subset$\textrm{Int}$\left(
\gamma\right)  $ is immaterial. If, on the other hand, we consider the case
$\left\vert \Gamma\right\vert \leq\ln N,$ then for large $N$ we can use the
relations $\left(  \ref{51}-\ref{53}\right)  $ for the contour $\gamma$ over
the landscape made by the union of the $x$-axis and the contour $\Gamma$ of
height $\leq\ln N.$ This is possible since $N^{1/2}\gg\ln N.$ So, as in
\textbf{1}, we conclude that $g^{\prime}(\omega^{-,k},N)\sim\frac{1}{N}.$

\textbf{4.} The general case of several $\left(  -\right)  $-segments follows
from the combinations of the three cases above.

\section{Conclusions and further comments}

We showed that the Schonmann projection of the extremal low-temperature Gibbs
measures of the 2-dimensional zero-field Ising model changes weakly at the
origin if conditioned on a long segment far from the origin, no matter how
long the segment is. The property responsible for this is a lack of entropic
repulsion from a long segment \textbf{in the direction of the segment}. We use
it to show that this projection is a $g$-measure, i.e. it has a kind of
one-sided Gibbsian property.

An earlier example of a non-Gibbsian $g$-measure was found in \cite{FGM}. On
the other hand, in \cite{BEEL} a Gibbsian non-$g$-measure is displayed. The
$g$-measure property thus cannot be seen as either weaker or stronger than the
property of being a Gibbs measure.

Presumably our result remains true for all subcritical temperatures, by
applying a coarse-graining argument as has been developed by Ioffe, Velenik
and their collaborators on Ornstein-Zernike behavior, see e.g. \cite{IOVW, OV}.

The $g$-function, although continuous, cannot be too regular, as the Schonmann
projection is known to be non-Gibbsian. It therefore cannot have the property
of "summable variations" (otherwise known as "Dini continuity"), as was
remarked before in \cite{BC}, and as also follows from \cite{BFV}. One can
wonder about "how continuous" or "how regular" the $g$-function might be. Our
proof suggests the following answer. If we consider the magnetization at the
origin $(0,0)$ conditioned on a segment $[\left(  -n,0\right)  ,\left(
-1,0\right)  ]$ to be all plus, the upper part of the enforced contour looks
like a (Brownian) bridge, of length $n$, constricted to be positive. Our
relations $\left(  \ref{51}-\ref{53}\right)  $ allow us to conclude that
\emph{the $n$-variation} $var_{n}(g)$ of $g$ at the "all-plus" configuration
is $O(\frac{1}{n})$, which is non-summable. An early mentioning of such an
argument is given in \cite{BF}, section VII. Notice again that the "all-plus"
configuration which is responsible for this behavior, is in some sense the
worst one, and is \emph{atypical} for the measure $\mu^{-}$ under
consideration. Indeed, for $\mu^{-}$-most configurations $\omega$, there
exists a positive density of minuses, and changing $\omega$ at a distance
larger than $n$, left of the origin, will only have an exponentially small
effect at site $(0,0)$. Incidentally, in \cite{BC} it was shown that for
\emph{typical} configurations a square summability condition is satisfied.

Although higher-dimensional versions of the non-Gibbsianness of the Schonmann
projected measures have been proved, \cite{MMR}, due to a similar entropic
repulsion argument there seems to be no natural higher-dimensional extension
of the $g$-measure property. One possible interpretation of a one-sided
conditioning would be requiring the Global Markov property, another one would
be requiring a continuous dependence on the lexicographic past. However, in
both these situations, there are counterexamples, or nearest-neighbour Gibbs
measure, thus having the Local Markov property, but lacking the Global Markov
property, or having conditional expectations which are discontinuous as a
function of the lexicographic past. For a discussion on some of those and
related issues, see e.g. \cite{ELP}.

\end{document}